\documentclass{jpsj3}
\usepackage{txfonts}
\usepackage{graphicx}
\usepackage{epstopdf}
\usepackage{dcolumn}
\usepackage{bm}
\usepackage{amsmath}
\usepackage{color}

\def\lesssim{\ \raise.3ex\hbox{$<$}\kern-0.8em\lower.7ex\hbox{$\sim$}\ }
\def\gesim{\ \raise.3ex\hbox{$>$}\kern-0.8em\lower.7ex\hbox{$\sim$}\ }

\title{Normal-state Properties of a Unitary Bose-Fermi Mixture: A Combined Strong-coupling Approach with Universal Thermodynamics}

\author{Digvijay Kharga\thanks{digvijay@rk.phys.keio.ac.jp}, Hiroyuki Tajima, Pieter van Wyk, Daisuke Inotani, and Yoji Ohashi}
\inst{Department of Physics, Keio University, 3-14-1 Hiyoshi, Kohuku-ku, Yokohama 223-8522, Japan} 

\abst{We theoretically investigate normal-state properties of a unitary Bose-Fermi mixture. Including strong hetero-pairing fluctuations,  we evaluate the Bose and Fermi chemical potential, internal energy, pressure, entropy, as well as specific heat at constant volume $C_V$, within the framework of a combined strong-coupling theory with exact thermodynamic identities. We show that hetero-pairing fluctuations at the unitarity cause non-monotonic temperature dependence of $C_V$, being qualitatively different from the monotonic behavior of this quantity in the weak- and strong-coupling limit. On the other hand, such an anomalous behavior is not seen in the other quantities. Our results indicate that the specific heat $C_V$, which has recently become observable in cold atom physics, is a useful quantity for understanding strong-coupling aspects of this quantum system.}

\begin{document}
\maketitle
\par
\section{Introduction}
\par
A tunable interaction associated with a Feshbach resonance is an advantage of cold atom physics\cite{Pethick,Chin}. In particular, this unique technique has extensively been used in cold Fermi gas physics\cite{Jin,Zwierlein,Bartenstein,Kinast}, for the study of BCS (Bardeen-Cooper-Schrieffer)-BEC (Bose-Einstein condensation) crossover phenomenon\cite{NSR,Randeria,Strinati,Ohashi,Chen,Bloch,Giorgini}. However, Feshbach resonances have also been discovered in various Bose-Fermi mixtures\cite{Stan,Inouye,Salomon,Ferlaino,Zaccanti,Deh,Schuster,Repp}. Using a hetero-nuclear Feshbach resonance, the formation of $^{40}$K-$^{87}$Rb molecules has also been observed\cite{Bongs2}. Since experimental efforts in cold Fermi gas physics have developed various techniques to measure physical quantities in strongly interacting atomic gases\cite{Stewart2008,Gaebler2010,Nascimbene2,Horikoshi2010,Sanner,Sommer,Esslinger,Ku,Lee,Sagi2015,Horikoshi}, it seems timely to extend the strong-coupling physics developed in cold Fermi gas physics to Bose-Fermi mixtures. Indeed, this approach has been started by several groups\cite{Yabu2,Yabu3,Yabu1,Suzuki1,Ref3,Ref4,Ref5,My,Vijay}.
\par
In this paper, we theoretically investigate normal-state properties of a unitary gas mixture of single-component Bose and Fermi atoms, as a typical strongly interacting Bose-Fermi mixture. Including Bose-Fermi hetero-pairing fluctuations associated with a tunable inter-species interaction, we examine strong-coupling corrections to thermodynamic quantities in the unitarity limit. We also clarify what observable quantity is sensitive to pairing fluctuations.  
\par
For later convenience, we explain two keys of our strategy here. First, regarding how to treat hetero-pairing fluctuations, Refs.\cite{My,Vijay} have pointed out that the ordinary non self-consistent $T$-matrix approximation (TMA)\cite{Strinati,Tsuchiya2009,Hui2010}, which has frequently been used for the study of ultracold Fermi gases in the BCS-BEC crossover region, has room for improvement, when it is applied to a Bose-Fermi mixture. Roughly speaking, this problem comes from the fact that TMA uses a {\it bare} Bose Green's function in evaluating self-energy corrections, so that the required gapless Bose excitations\cite{Pines} are not realized there at the BEC phase transition temperature $T_{\rm BEC}$. To cure this, Ref.\cite{My} improved TMA so that this condition can be satisfied in the Bose Green's function used in the self-energy. In this paper, we employ this improved TMA (iTMA). 
\par
To explain the second key, we note that the validity of an approximate strong-coupling theory usually depends on the physical quantity that we consider. In cold Fermi gas physics, for example, while the Gaussian pair-fluctuation theory\cite{NSR} can describe the BCS-BEC crossover behavior of the superfluid phase transition temperature $T_{\rm c}$, it unphysically gives negative single-particle density of states in the unitarity regime\cite{Tsuchiya2009}. Although this problem is overcome in TMA\cite{Tsuchiya2009}, it unphysically gives negative spin susceptibility $\chi$ in the strong-coupling regime\cite{Kashimura}. Thus, when one examines various physical quantities within a strong-coupling theory, the agreement of a calculated result with an experimental result does not guarantee the correctness of the other results. 
\par
To overcome this problem, some of the authors have recently proposed to combine a strong-coupling theory with exact thermodynamic identities\cite{Tajima}. In this proposal, complicated (and approximate) strong-coupling calculations are only executed to obtain a thermodynamic quantity ($\equiv X$). Other quantities are then evaluated from $X$ by using exact thermodynamic relations, instead of repeating strong-coupling calculations. The advantage of this approach is that all the calculated quantities are related to one another through {\it exact} thermodynamic identities. Thus, when one of them is checked by an experiment, the correctness of the others is simultaneously confirmed. On the other hand, when a calculated quantity disagrees with an experimental result, one may improve the strong-coupling theory, to repeat this approach. This idea has been applied to a superfluid $^6$Li Fermi gas\cite{Tajima}, to successfully explain various observed quantities. In this paper, we applied this approach to a unitary Bose-Fermi mixture. In particular, at the unitary, we can conveniently use the so-called universal thermodynamics developed by Ho\cite{Ho1}. In this sense, our approach may be called ``combined iTMA with universal thermodynamics."
\par
This paper is organized as follows: In Sec.2, we explain our formulation. We first explain how to calculate the pressure, entropy, internal energy, as well as specific heat, by using the thermodynamic identities in the unitarity limit, for given Bose ($\mu_{\rm B}$) and Fermi ($\mu_{\rm F}$) chemical potential. We then present iTMA to evaluate $\mu_{{\rm s}={\rm B,F}}$. We show calculated thermodynamic quantities in a unitary Bose-Fermi mixture in Sec.3, which is followed by the summary in Sec. 4. Throughout this paper we take $\hbar=k_{\rm B}=1$, for simplicity.
\par
\section{Formulation}
\par
We consider a uniform Bose-Fermi mixture, consisting of single-component Bose and Fermi atoms, described by the model Hamiltonian\cite{Ref3,My},
\begin{eqnarray}
H&=&\sum_{{\bm p},{\rm s=B,F}}
\xi_{\bm p}^{\rm s}
c_{{\bm p},{\rm s}}^{\dagger}c_{{\bm p},{\rm s}}
\nonumber
\\
&-&
U_{\rm BF}\sum_{{\bm p},{\bm p}',{\bm q}}
c_{{\bm p}+{\bm q}/2,{\rm B}}^\dagger
c_{-{\bm p}+{\bm q}/2,{\rm F}}^{\dagger}
c_{-{\bm p}'+{\bm q}/2,{\rm F}}
c_{{\bm p}'+{\bm q}/2,{\rm B}},
\label{eq.1}
\end{eqnarray}
where $c_{{\bm p},{\rm s}={\rm B,F}}^\dagger$ is the creation operator of a Bose (s=B) and Fermi (s=F) atom, with the kinetic energy $\xi_{\bm p}^{\rm s}=\varepsilon_{\bm p}-\mu_{\rm s}={\bm p}^2/(2m)-\mu_{\rm s}$, measured from the chemical potential $\mu_{\rm s}$. Here, we have assumed the mass-balanced case, $m_{\rm B}=m_{\rm F}\equiv m$, for simplicity. $-U_{\rm BF}$ ($<0$) is an inter-species attractive interaction associated with a hetero-Feshbach resonance. The unitary Bose-Fermi mixture is characterized by the vanishing inverse $s$-wave scattering length $a_{\rm BF}^{-1}=0$, where $a_{\rm BF}$ is related to $-U_{\rm BF}$ as,\begin{equation}
{4\pi a_{\rm BF} \over m}=-
{U_{\rm BF} \over 
\displaystyle 
1-U_{\rm BF}\sum_{\bm p}^{p_{\rm c}}
{1 \over 2\varepsilon_{\bm p}}},
\label{eq.0}
\end{equation} 
with $p_{\rm c}$ being a high-momentum cutoff.
\par
This paper deals with the simple case when the number $N_{\rm F}$ of Fermi atoms equals the number $N_{\rm B}$ of Bose atoms ($N_{\rm F}=N_{\rm B}\equiv N$). In this case, while the system in weak-coupling limit is an ideal gas mixture of $N$ bosons and $N$ fermions, the strong-coupling limit is described as an ideal gas of $N$ composite Fermi molecules. We also ignore effects of a harmonic trap, as well as intra-species interactions, for simplicity. Inclusion of these remains as our future problem. 
\par
The vanishing inverse scattering length $a_{\rm BF}^{-1}=0$ at the unitarity simplifies the thermodynamics, which is sometimes referred to as the universal thermodynamics in the literature\cite{Ho1}. Since the energy scale associated with the inter-species interaction no longer exists, the relevant energy scales are only $T$ and $\mu_{{\rm s}={\rm B,F}}$, as in the non-interacting case. Then, the thermodynamic potential $\Omega$ can be expressed as
\begin{equation}
\Omega={TV \over \lambda_T^3}F(X_{\rm B},X_{\rm F}),
\label{eq.2}
\end{equation}
where $V$ is the system volume, and $\lambda_T=\sqrt{2\pi/(mT)}$ is the thermal de-Broglie wavelength. $F(X_{\rm B},X_{\rm F})$ is a dimensionless function involving the dimensionless variables $X_{{\rm s}={\rm B,F}}=\mu_{\rm s}/T$. We emphasize that this simple scaling form is not satisfied, when $a_{\rm BF}^{-1}\ne 0$ (except in the trivial non-interacting case). Using the thermodynamic identity, 
\begin{equation}
N_{{\rm s}={\rm B,F}}=
-\left(
{\partial\Omega \over \partial\mu_{\rm s}}
\right)_{T,V,\mu_{\rm -s}}
=-{V \over \lambda_T^3}
\left(
{\partial F \over \partial X_{\rm s}}
\right)_{X_{\rm -s}}
\label{eq.3}
\end{equation}
(where ``-s" means the opposite component to ``s"), we obtain
\begin{eqnarray}
dF(X_{\rm B},X_{\rm F})
&=&\sum_{{\rm s}={\rm B,F}}
\left(
{\partial F(X_{\rm B},X_{\rm F}) \over \partial X_{\rm s}}
\right)_{X_{-{\rm s}}}dX_{\rm s}
\nonumber
\\
&=&
-\sum_{{\rm s}={\rm B,F}}
{N_{\rm s}\lambda_T^3 \over TV}
\left[
{\partial \mu_{\rm s}(T) \over \partial T}-{\mu_{\rm s}(T) \over T}
\right]
dT.
\label{eq.3b}
\end{eqnarray}
Noting that $N_{\rm B}=N_{\rm F}=V(2m\varepsilon_{\rm F})^{3/2}/(6\pi^2)$ in the present case (where $\varepsilon_{\rm F}$ is the atomic Fermi energy), we can evaluate $F(X_{\rm B}, X_{\rm F})$ from the equation,
\begin{equation}
F(X_{\rm B},X_{\rm F})
=
-{4\varepsilon_{\rm F}^{3/2} \over 3\sqrt{\pi}}
\sum_{{\rm s}={\rm B,F}}
\int_{\infty}^{T}
{dT' \over T'^{5/2}}
\left[
{d\mu_{\rm s}(T') \over dT'}-{\mu_{\rm s}(T') \over T'}
\right]
.
\label{eq.4}
\end{equation}
In obtaining Eq. (\ref{eq.4}), we have used the fact that $F(X_{\rm B},X_{\rm F})$ vanishes when $T\to\infty$, because the system is reduced to a classical ideal gas ($\Omega\propto T$) there. 
\par
Once $F(X_{\rm B},X_{\rm F})$ is determined, the pressure $P$, entropy $S$, and the internal energy $E$, are obtained from the identities,
\begin{eqnarray}
\left\{
\begin{array}{l}
\displaystyle
P=-\Omega/V=-{T \over \lambda_T^3}F(X_{\rm B},X_{\rm F}),\\
\displaystyle
S=-\left({\partial \Omega \over \partial T}\right)_{V,\mu_{\rm s}}=
V\left({\partial P \over \partial T}\right)_{\mu_{\rm s}},\\
\displaystyle
E=-PV+TS+\sum_{{\rm s}={\rm B,F}}\mu_{\rm s}N_{\rm s}={3 \over 2}PV.
\end{array}
\right.
\label{eq.5}
\end{eqnarray}
Although the last expression in the third line in Eq. (\ref{eq.5}) is the same as the well-known formula in an ideal gas, it is a result of the universal thermodynamics for a unitary gas. Indeed, because the expression for the entropy $S$ in Eq. (\ref{eq.5}) can be calculated to be
\begin{eqnarray}
S
&=&-{5V \over 2\lambda_T^3}F(X_{\rm B},X_{\rm F})
-{TV \over \lambda_T^3}\sum_{{\rm s}={\rm B,F}}
\left(
{\partial F(X_{\rm B},X_{\rm F}) \over \partial X_{\rm s}}
\right)_{X_{-{\rm s}}}
\left(
{\partial X_{\rm s} \over \partial T}
\right)_{\mu_{\rm s}}
\nonumber
\\
&=&
{5 \over 2}{PV \over T}-{1 \over T}
\sum_{{\rm s}={\rm B,F}}N_{\rm s}\mu_{\rm s},
\label{eq.5a}
\end{eqnarray}
this relation is immediately confirmed by substituting Eq. (\ref{eq.5a}) into the first expression in the third line on Eq. (\ref{eq.5}). Because this simple relation holds in the present unitary Bose-Fermi mixture we are considering, we actually do not have to calculate the entropy $S$, in evaluating the specific heat at constant volume $C_V$. That is, $C_V$ can be obtained from
\begin{equation}
C_V=
\left(
{\partial E \over \partial T}
\right)_V
=\frac{3}{2}V{dP \over dT}
\label{eq.5b}.
\end{equation}
\par
\begin{figure}[t]
\begin{center}
\includegraphics[width=12cm]{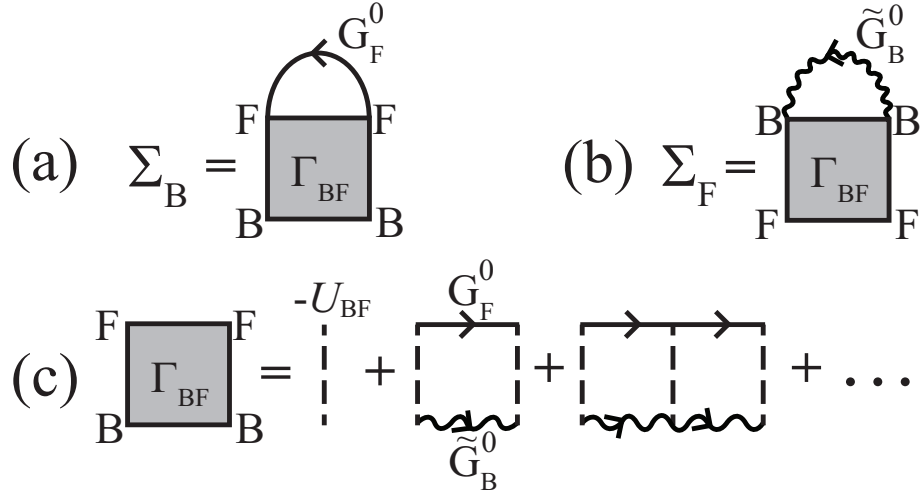}
\caption{Self-energies in iTMA. (a) Bose component $\Sigma_{\rm B}({\bm p},i\omega_n^{\rm B})$. (b) Fermi component $\Sigma_{\rm F}({\bm p},i\omega_n^{\rm F})$. (c) Bose-Fermi scattering matrix $\Gamma_{\rm BF}$. The solid line is the bare single-particle Fermi Green's function $G_{\rm F}^0=[i\omega_n^{\rm F}-\xi_{\bm p}^{\rm F}]^{-1}$. The wavy line represents the Bose Green's function ${\tilde G}_{\rm B}^0$ in Eq. (\ref{eq.9}). The dashed line denotes the inter-species pairing interaction $-U_{\rm BF}~(<0)$.}
\label{fig1}
\end{center}
\end{figure}
\par
In our combined iTMA approach with universal thermodynamics, the remaining is to determine $\mu_{{\rm s}={\rm B,F}}(T)$ within the framework of iTMA\cite{My}. In this strong-coupling theory, these are determined from the coupled number equations for bosons and fermions, 
\begin{equation}
N_{{\rm s}={\rm B,F}}=\mp T\sum_{{\bm p},\omega_n^{\rm s}}
G_{\rm s}({\bm p},i\omega_n^{\rm s}).
\label{eq.6}
\end{equation}
Here, we take the upper (lower) sign for the Bose (Fermi) component in this paper. $\omega_n^{\rm B}$ and $\omega_n^{\rm F}$ are boson and fermion Matsubara frequency, respectively. In these number equations, effects of hetero-pairing fluctuations are described by the self-energies $\Sigma_{{\rm s}={\rm B,F}}({\bm p},i\omega_n^{\rm s})$ in the single-particle Bose ($G_{\rm B}$) and Fermi ($G_{\rm F}$) thermal Green's function,
\begin{equation}
G_{{\rm s}={\rm B,F}}({\bm p},i\omega_n^{\rm s})=
{1 
\over 
i\omega_n^{\rm s}-\xi_{\bm p}^{\rm s}-\Sigma_{\rm s}({\bm p},i\omega_n^{\rm s})
}.
\label{eq.8}
\end{equation}
As shown in Fig.\ref{fig1}, the iTMA self-energies have the same diagrammatic structures as in the case of the ordinary non self-consistent $T$-matrix approximation (TMA)\cite{Ref3}. One exception is that the bare Bose Green's function $G_{\rm B}^0({\bm p},i\omega_n^{\rm B})=[i\omega_n^{\rm B}-(\varepsilon_{\bm p}-\mu_{\rm B})]^{-1}$ in TMA is replaced by\cite{My}
\begin{equation}
{\tilde G}^0_{\rm B}({\bm p},i\omega_n^{\rm B})=
{1 \over i\omega_n^{\rm B}-{\tilde \xi}_{\bm p}^{\rm B}}.
\label{eq.9}
\end{equation}
Here, ${\tilde \xi}_{\bm p}^{\rm B}=\varepsilon_{\bm p}-{\tilde \mu}_{\rm B}$, where the effective Bose chemical potential ${\tilde \mu}_{\rm B}=\mu_{\rm B}-\Sigma_{\rm B}(0,0)$ involves the self-energy correction at ${\bm p}=\omega_n^{\rm B}=0$. Regarding this modification, we note that $T_{\rm BEC}$ is determined from the condition that Bose excitations become gapless in the BEC phase below $T_{\rm BEC}$. In both TMA and iTMA, the dressed Bose Green's function $G_{\rm B}({\bm p},i\omega_n^{\rm B})$ in Eq. (\ref{eq.8}) satisfies this required condition, when
\begin{equation}
\mu_{\rm B}=\Sigma_{\rm B}({\bm p}=0,i\omega_n^{\rm B}=0).
\label{eq.HP}
\end{equation}
Equation (\ref{eq.HP}) may be viewed as an extension of the Hugenholtz-Pines condition\cite{Pines} for interacting bosons to a Bose-Fermi mixture. (We explain this extension in Appendix A.) On the other hand, the bare Bose Green's function in TMA still exhibits gapped excitations even when Eq. (\ref{eq.HP}) is satisfied at $T_{\rm BEC}$, leading to the underestimation of low-energy Bose excitations in the TMA self-energies $\Sigma_{{\rm s}={\rm B,F}}({\bm p},i\omega_n^{\rm s})$ near $T_{\rm BEC}$. Although iTMA is still a non self-consistent approximation, the required gapless Bose excitations at $T_{\rm BEC}$ are correctly taken into account, because the Bose Green's function in Eq. (\ref{eq.9}) is used everywhere in $\Sigma_{{\rm s}={\rm B,F}}({\bm p},i\omega_n^{\rm s})$. (For more details about iTMA, we refer to Ref.\cite{My}.)
\par
Summing up the diagrams in Fig. \ref{fig1}, we obtain\cite{My,Vijay}
\begin{equation}
\Sigma_{{\rm s}={\rm B,F}}({\bm p}, i\omega_n^{\rm s})=\pm
T\sum_{{\bm q},\omega_{n'}^{\rm F}}
\Gamma_{\rm BF}({\bm q},i\omega_{n'}^{\rm F})
G_{\rm -s}^0({\bm q}-{\bm p},i\omega_{n'}^{\rm F}-i\omega_{n}^{\rm s}),
\label{eq.11}
\end{equation}
where
\begin{eqnarray}
\Gamma_{\rm BF}({\bm q},i\omega_n^{\rm F})
&=&
-{U_{\rm BF} \over 1-U_{\rm BF}\Pi_{\rm BF}({\bm q},i\omega_n^{\rm F})}
\nonumber
\\
&=&
{1 
\over 
\displaystyle
{m \over 4\pi a_{\rm BF}}+
\left[
\Pi_{\rm BF}({\bm q},i\omega_n^{\rm F})
-
\sum_{\bm p}^{p_{\rm c}}{1 \over 2\varepsilon_{\bm p}}
\right]
}
\nonumber
\\
&=&
{1 
\over
\Pi_{\rm BF}({\bm q},i\omega_n^{\rm F})
-\sum_{\bm p}^{p_{\rm c}}{1 \over 2\varepsilon_{\bm p}}
}
\label{eq.13}
\end{eqnarray}
is the iTMA Bose-Fermi scattering matrix, describing effects of hetero-pairing fluctuations. In obtaining the last expression in Eq. (\ref{eq.13}), we have taken the unitarity limit $a_{\rm BF}^{-1}\to 0$. In Eq. (\ref{eq.13}),
\begin{eqnarray}
\Pi_{\rm BF}({\bm q},i\omega_n^{\rm F})
&=&
T\sum_{{\bm k},i\omega_{n'}^{\rm B}}
G_{\rm F}^0({\bm q}-{\bm k},i\omega_n^{\rm F}-i\omega_{n'}^{\rm B})
{\tilde G}_{\rm B}^0({\bm k},i\omega_{n'}^{\rm B})
\nonumber
\\
&=&
-\sum_{\bm k}
{1-f_{\rm F}(\xi_{{\bm k}+{\bm q}/2}^{\rm F})
+f_{\rm B}(\tilde{\xi}_{-{\bm k}+{\bm q}/2}^{\rm B})
\over
i\omega_n^{\rm F}
-\xi_{{\bm k}+{\bm q}/2}^{\rm F}-{\tilde \xi}_{-{\bm k}+{\bm q}/2}^{\rm B}}
\label{eq.14}
\end{eqnarray}
is the lowest-order hetero-pair correlation function, where $f_{\rm s}(x)$ is the Bose (s=B) and Fermi (s=F) distribution function. In the first line in Eq. (\ref{eq.14}), $G_{\rm F}^0=[i\omega_n^{\rm F}-\xi_{\bm p}^{\rm F}]^{-1}$ is the bare fermion Green's function.
\par
\begin{figure}
\begin{center}
\includegraphics[width=9cm]{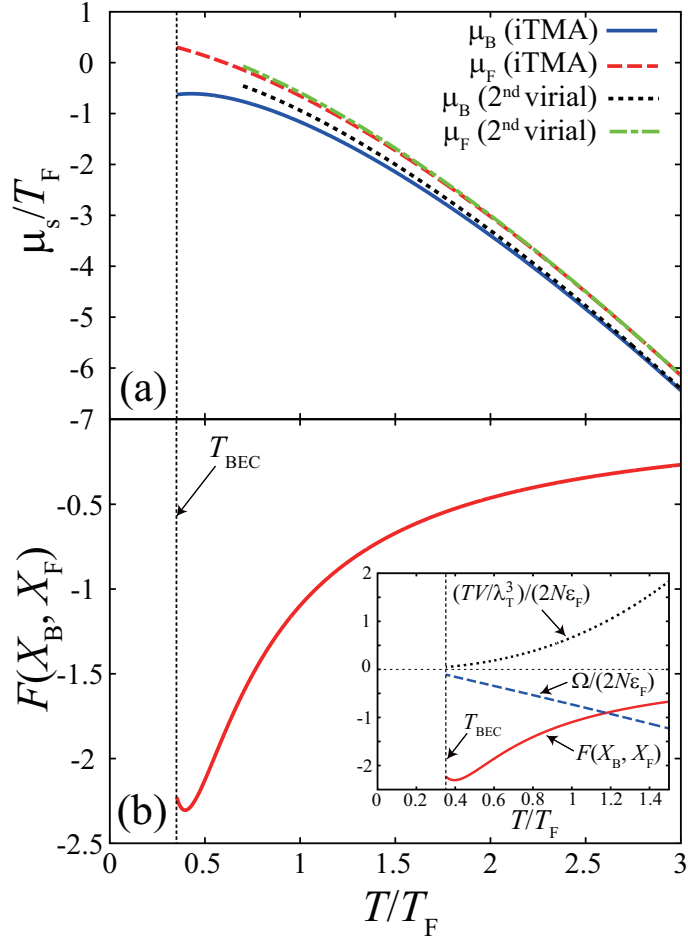}
\caption{(Color online) (a) Calculated Bose ($\mu_{\rm B}$) and Fermi ($\mu_{\rm F}$) chemical potential in a unitary Bose-Fermi mixture above $T_{\rm BEC}$ ($=0.35T_{\rm F}$, where $T_{\rm F}$ is the Fermi temperature of an $N$ Fermi atomic gas). In this figure, ``$2^{\rm nd}$ virial" shows the result by the second-order virial expansion method. (b) The scaling function $F(X_{\rm B},X_{\rm F})$ as a function of temperature, which is calculated from Eq. (\ref{eq.4}). The inset shows the thermodynamic potential $\Omega$, as well as the factor $VT/\lambda_T^3$ appearing in Eq. (\ref{eq.2}).}
\label{fig2}
\end{center}
\end{figure}
\par
\section{Thermodynamic properties of a unitary Bose-Fermi mixture}
\par
Figure \ref{fig2}(a) shows the calculated chemical potential $\mu_{{\rm s}={\rm B,F}}$ in a unitary Bose-Fermi mixture above $T_{\rm BEC}$. To evaluate $F(X_{\rm B},X_{\rm F})$ from Eq. (\ref{eq.4}), one needs $\mu_{{\rm s}={\rm B,F}}$ up to the high-temperature region ($T/T_{\rm F}\gg 1$). However, we actually do not have to use iTMA to such the classical region, because Fig. \ref{fig2} shows that the iTMA results are well reproduced by the second-order virial expansion\cite{Dashen,Huang,Brown,Ho2,Liu,Kaplan,Virial,Parish}, when $T/T_{\rm F}\gesim 3$ (where $T_{\rm F}$ is the Fermi temperature of $N$ Fermi atoms). In this high-temperature expansion, the number equations are given by,
\begin{equation}
N_{\rm B}=
{V \over \lambda_{\rm T}^3}
\left[
B_{01}z_{\rm B}+2B_{02}z_{\rm B}^2+B_{11}z_{\rm F}z_{\rm B}
\right],
\label{eq.15}
\end{equation}
\begin{equation}
N_{\rm F}=
{V \over \lambda_{\rm T}^3}
\left[
B_{10}z_{\rm F}+2B_{20}z_{\rm F}^2+B_{11}z_{\rm F}z_{\rm B}
\right].
\label{eq.16}
\end{equation}
where $z_{\rm s}=e^{\mu_{\rm s}/T}$ is the fugacity, and the virial coefficients are given as $B_{n0}=(-1)^{n+1}n^{-5/2}$, $B_{0n}=n^{-5/2}$, and $B_{11}=\sqrt{2}$. (We summarize the derivation of Eqs. (\ref{eq.15}) and (\ref{eq.16}) in Appendix B.) Thus, we may only use iTMA in the low temperature region ($T/T_{\rm F}\lesssim 3$). For the high temperature classical region, we use the results obtained from the second-order virial expansion. Using this prescription, we numerically calculate the scaling function $F(X_{\rm B},X_{\rm F})$ from  Eq. (\ref{eq.4}), which gives Fig. \ref{fig2}(b). We briefly note that, although $F(X_{\rm B},X_{\rm F})$ in Fig. \ref{fig2}(b) exhibits a non-monotonic behavior near $T_{\rm BEC}$, the thermodynamic potential $\Omega$, which is given by the product of $F(X_{\rm B},X_{\rm F})$ and $VT/\lambda_T^3~(\propto T^{5/2})$ (see Eq. (\ref{eq.2})), monotonically depends on the temperature. To explicitly show this, we plot $\Omega$ as a function of temperature in the inset in Fig. \ref{fig2}(b).
\par
\begin{figure}
\begin{center}
\label{fig:fig3}
\includegraphics[width=8cm]{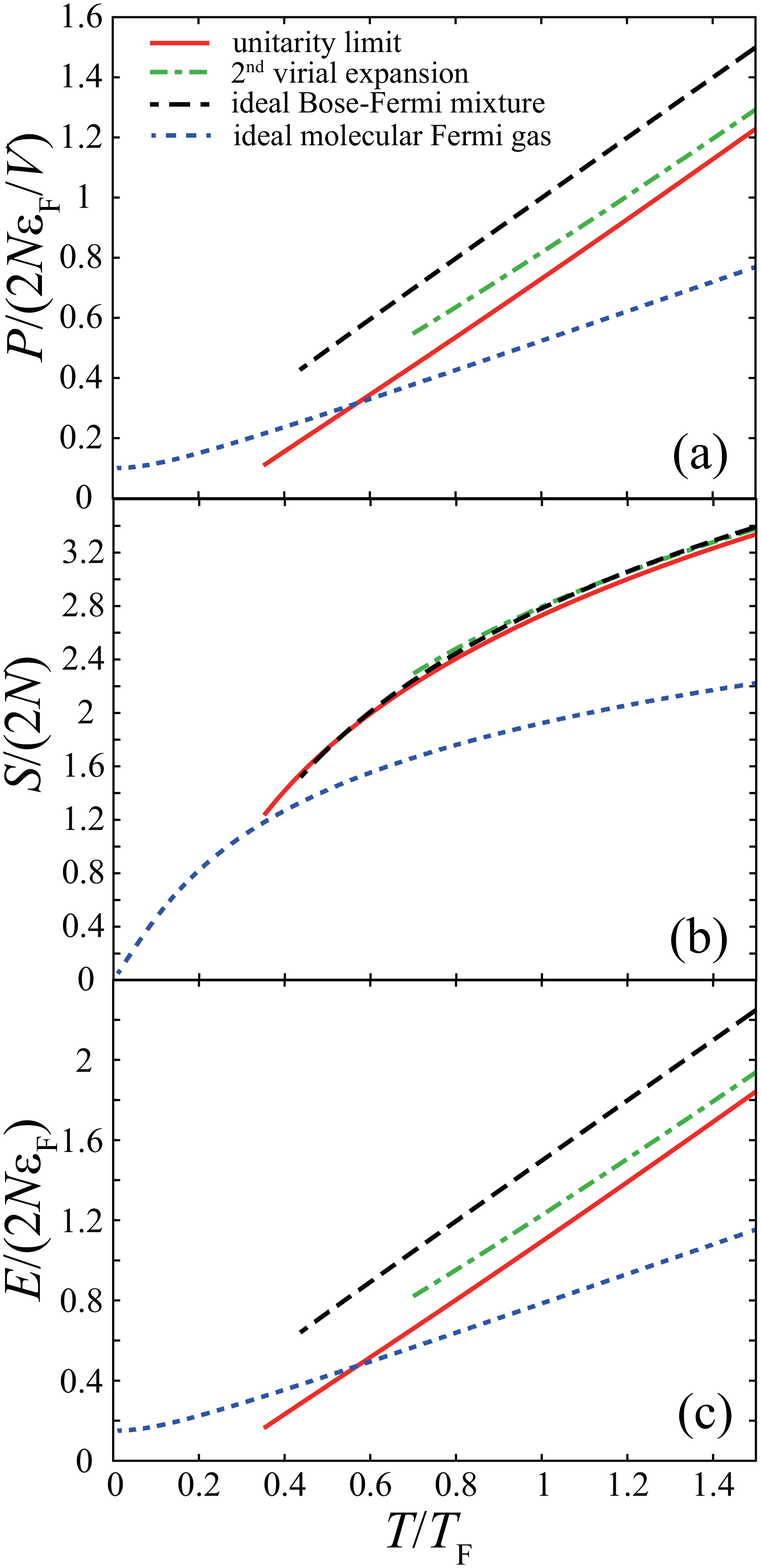}
\caption{(Color online) Calculated thermodynamic quantities. (a) pressure P. (b) entropy S. (c) internal energy E. In panel (c), the contribution from the molecular binding energy $E_{\rm bind}=-N/(ma_{\rm BF}^2)$ is subtracted from $E$. In this figure, as well as in Fig. \ref{fig4}, ``$2^{\rm nd}$ virial expansion" shows the result when $\mu_{{\rm s}={\rm B,F}}$ obtained from the coupled number equations (\ref{eq.15}) and (\ref{eq.16}) are used in Eq. (\ref{eq.4}) down to the low temperature region. For comparison, each panel also shows the result in weak-coupling limit (``ideal Bose-Fermi mixture," where $T_{\rm BEC}=0.44T_{\rm F}$), as well as that in the strong-coupling limit (``ideal molecular Fermi gas," where BEC does not occur).}
\label{fig3}
\end{center}
\end{figure}
\par
\begin{figure}
\begin{center}
\label{fig:fig6}
\includegraphics[width=9cm]{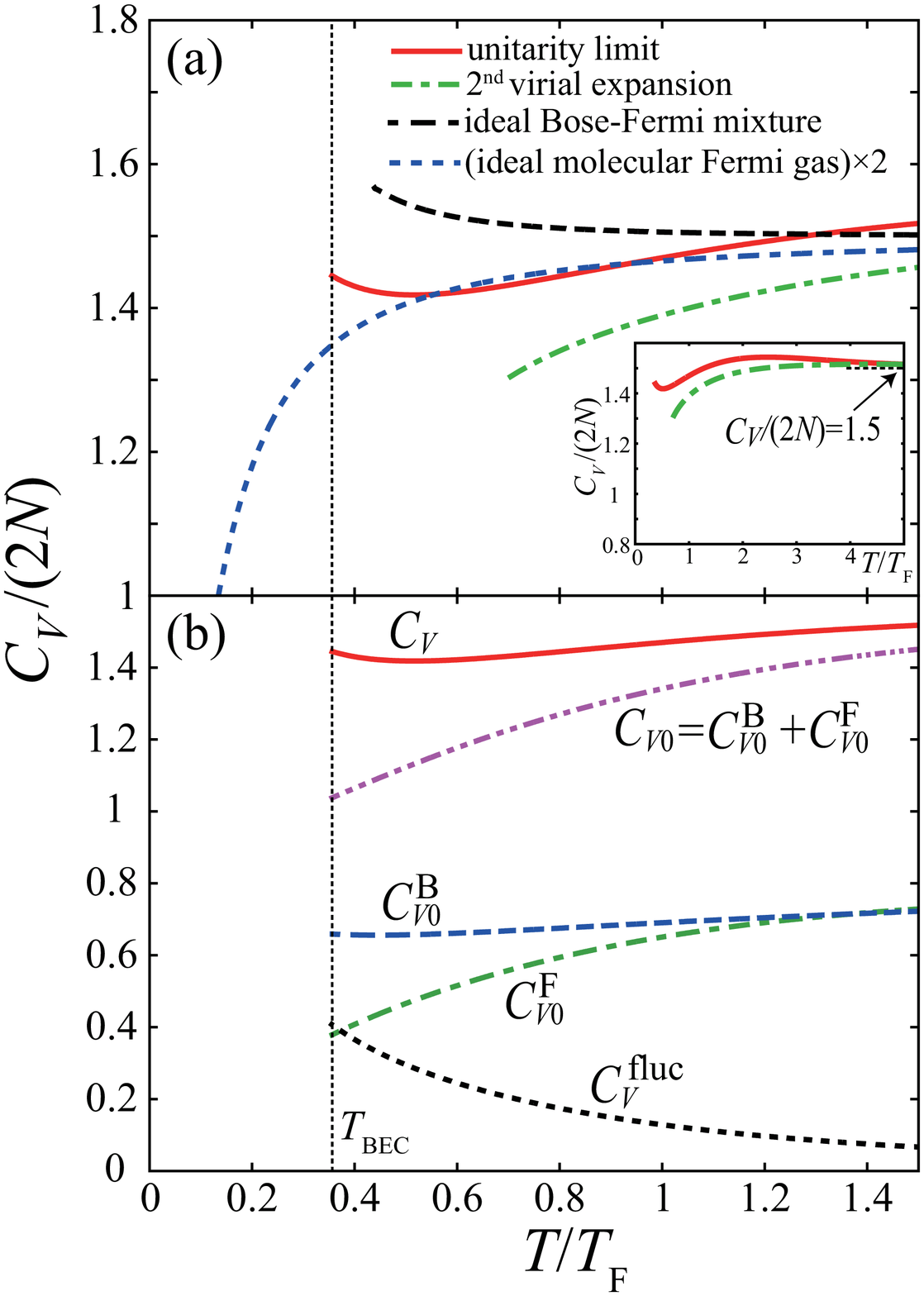}
\caption{(Color online) (a) Calculated specific heat at constant volume $C_V$ in a unitary Bose-Fermi mixture. The inset shows the result to the high-temperature region, where $C_V$ is found to approach the expected value $C_V/(2N)=1.5$. In panel (b), we decompose the specific heat $C_V$ into the sum of (1) free Bose atomic gas component $C_{V0}^{\rm B}$, (2) free Fermi atomic gas component $C_{V0}^{\rm F}$, and (3) fluctuation contribution $C_V^{\rm fluc}\equiv C_V-C_{V0}^{\rm B}-C_{V0}^{\rm F}$.}
\label{fig4}
\end{center}
\end{figure}
\par
Once the function $F(X_{\rm B},X_{\rm F})$ is determined, we can straightforwardly calculate $P$, $S$, $E$, and $C_V$, from the thermodynamic identities in Eqs. (\ref{eq.5}), and (\ref{eq.5b}). 
\par
Figure \ref{fig3} shows the calculated thermodynamic quantities in Eq. (\ref{eq.5}). Although the unitarity limit ($a_{\rm BF}^{-1}=0$) is located between the weak-coupling limit ($a_{\rm BF}^{-1}=-\infty$) and the strong-coupling limit ($a_{\rm BF}^{-1}=+\infty$), their temperature dependence is rather similar to the former case. This is because a two-body bound state is only possible when $a_{\rm BF}^{-1}\ge 0$ (with the binding energy $E_{\rm bind}=-1/(ma_{\rm BF}^2)$). Thus, although hetero-pairing fluctuations are enhanced near $T_{\rm BEC}$, they still do not acquire a clear molecular character in a unitary Bose-Fermi mixture.
\par
We note that the difference between the internal energy $E_{\rm unitarity}$ of the present unitary Bose-Fermi mixture and the internal energy $E_{\rm free}$ of an ideal Bose-Fermi mixture seen in Fig. \ref{fig3}(c) comes from interaction effects (although the interaction parameter $a_{\rm BF}^{-1}$ formally vanishes in the former). When we write the former energy as $E_{\rm unitarity}=\eta_{\rm BF}E_{\rm free}$, one finds, for example at $T_{\rm BEC}$ of the ideal Bose-Fermi mixture,
\begin{equation}
\eta_{\rm BF}=0.444.
\label{eq.z}
\end{equation}
The measurement of this parameter would be useful for the estimation of interaction effects in a unitary Bose-Fermi mixture, as well as the assessment of the present theoretical approach. We briefly note that, because the internal energy is directly related to the pressure Fig. \ref{fig3}(a) (see the third line in Eq. (\ref{eq.4})), one can also evaluate this parameter $\eta_{\rm BF}$ from the observation of the pressure $P$, by using the combined Gibbs-Duhem relation with the local density approximation developed in Ref.\cite{Ho2010}.
\par
However, apart from such quantitative comparison, we also note that all the quantities shown in Fig. \ref{fig3} monotonically decrease with decreasing the temperature, irrespective of the interaction strength. In addition, in each case, the temperature dependence is not so different from the result by the virial expansion. Thus, although these quantities, $P$, $S$, and $E$, would be affected by hetero-pairing fluctuations, it seems difficult to extract detailed strong-coupling phenomena from the simple comparison with classical results (at $T/T_{\rm F}\gg 1$) and results in the weak- and strong-coupling limit. 
\par
In contrast, Fig. \ref{fig4}(a) shows that the specific heat at constant volume $C_V$ exhibits {\it non-monotonic} behavior at the unitarity, which is qualitatively different from the monotonic increase (decrease) with decreasing the temperature seen in the weak-coupling (strong-coupling) limit. In addition, while $C_V$ agrees with the result by the second-order virial-expansion at high temperatures (see the inset in Fig. \ref{fig4}), the non-monotonic temperature dependence cannot by explained by this high-temperature expansion, implying that quantum many-body effects play a crucial role for this anomaly.
\par   
To understand the anomalous behavior of $C_V$ in Fig. \ref{fig4}(a), it is convenient to examine the contribution of free Bose atoms $C_{V0}^{\rm B}$ and that of free Fermi atoms $C_{V0}^{\rm F}$, given by, respectively,
\begin{equation}
C_{V0}^{\rm B}=
{d \over dT}\sum_{\bm p}\varepsilon_{\bm p}
f_{\rm B}(\varepsilon_{\bm p}-{\tilde \mu}_{\rm B}),\\
\label{eq.17a}
\end{equation}
\begin{equation}
C_{V0}^{\rm F}=
{d \over dT}\sum_{\bm p}\varepsilon_{\bm p}
f_{\rm F}(\varepsilon_{\bm p}-\mu_{\rm F}),
\label{eq.17}
\end{equation}
where $\mu_{{\rm s}={\rm B,F}}$ in Fig. \ref{fig2} are used. In the ordinary ideal Bose gas, the specific heat increases with decreasing the temperature. However, Fig. \ref{fig4}(b) shows that $C_{V0}^{\rm B}$ does not exhibit such a behavior. This means that, although the inter-species interaction at the unitarity is not strong enough to form Fermi molecules, enhanced hetero-pairing fluctuations near $T_{\rm BEC}$ suppress the number of Bose atoms contributing to BEC. Thus, the up-turn behavior of $C_V$ around $T\lesssim 0.5T_{\rm F}$ in Fig. \ref{fig4}(a) cannot be explained by the ordinary enhancement of the specific heat known in an ideal Bose gas near $T_{\rm BEC}$. 
\par
Since the sum $C_{V0}=C_{V0}^{\rm B}+C_{V0}^{\rm F}$ monotonically decreases at low temperatures (see Fig. \ref{fig4}(b)), the anomalous behavior of $C_V$ at the unitarity is considered as a strong-coupling effect. That is, the specific heat $C_V$ exhibits a dip structure around the temperature ($\equiv T_{\rm dip}\simeq 0.5T_{\rm F}$) where the enhancement of the fluctuation contribution $C_V^{\rm fluc}$ exceeds the suppression of the free atomic contribution $C_{V0}$. In this sense, the dip temperature $T_{\rm dip}$ may be physically interpreted as a characteristic temperature below which hetero-pairing fluctuations are important.
\par
Before ending this section, we note that it is helpful for experiments to theoretically clarify how to distinguish between the upturn behavior of $C_V$ seen in an ideal Bose-Fermi mixture near $T_{\rm BEC}$ and that caused by hetero-pairing fluctuations. For this purpose, it would be useful to examine how the temperature dependence of $C_V$ varies, as one moves from the weak-coupling regime to the unitarity limit. In this case, if the upturn behavior of $C_V$ once vanishes at an interaction strength ($\equiv{\tilde a}_{\rm BF}^{-1}~(<0)$), the upturn behavior of $C_V$ in the region ${\tilde a}_{\rm BF}^{-1}\le a_{\rm BF}^{-1}\le 0$ can physically be regarded as effects of hetero-pairing fluctuations. Since the present theory uses the universal thermodynamics\cite{Ho1}, it is not applicable to the region away from the unitarity limit. To explore the above-mentioned possibility, we need to improve the present theory to that where the universal thermodynamics is not necessary, which remains as our future challenge.
\par
\section{Summary}
\par
To summarize, we have discussed normal state properties of a Bose-Fermi mixture in the unitarity limit. Within the framework of a combined improved $T$-matrix approximation (iTMA) with universal thermodynamics, we have calculated several thermodynamic quantities ($P$, $S$, $E$, and $C_V$). Among them, the specific heat at constant volume $C_V$ was found to be a useful quantity for the study of strong-coupling effects in this system, because hetero-pairing fluctuations cause an anomalous non-monotonic temperature dependence of $C_V$.
\par
In this paper, we have ignored some realistic situations, such as a harmonic trap, mass difference between Bose and Fermi atoms, and intra-species interactions. In the next step, inclusion of these would be important. In addition, assessment of the present theoretical approach, especially the validity of iTMA, also remains as our future challenge. To examine thermodynamic properties of a Bose-Fermi mixture in the whole interaction regime from the weak- to strong-coupling limit, we also need to go beyond the present approach using the universal thermodynamics (which is only valid for  a unitary gas). Since the specific heat has recently become possible to observe\cite{Ku}, our results would contribute to the further development of strong-coupling physics of a Bose-Fermi mixture.
\par
\par
\begin{acknowledgment}
We thank R. Hanai, and D. Kagamihara for discussions. This work was supported by KiPAS project at Keio University. DK thanks Japan International Co-operation Agency (JICA) and Keio Leading-edge Laboratory of Science and  Technology (KLL) at Keio University for supporting this research. HT was supported by Grant-in-Aid for JSPS fellows (No. JP17J03975). DI was supported by Grant-in-Aid for Young Scientists (B) (No. JP16K17773) from JSPS. YO was supported by Grand-in-Aid for Scientific Research from MEXT and JSPS in Japan (No. JP15K00178, No. JP15H00840, No. JP16K05503).
\end{acknowledgment}
\par
\par
\appendix
\section{Extension of Hugenholtz-Pines condition to a Bose-Fermi mixture}
\par
In this appendix, we show that the Hugenholtz-Pines condition\cite{Pines} for interacting bosons can be extended to a Bose-Fermi mixture. For this purpose, we conveniently consider the model Hamiltonian in Eq.(\ref{eq.1}) in the coordinate-space representation, 
\begin{eqnarray}
H&=&\sum_{{\rm s}={\rm B,F}}
\int d{\bm r} \psi_{\rm s}^\dagger({\bm r})
\left[-{{\nabla^2} \over {2m}}-\mu_{\rm s}
\right]
\psi_{\rm s}({\bm r})
\nonumber
\\
&-&
U_{\rm BF}
\int d{\bm r} \psi_{\rm B}^\dagger({\bm r})
\psi_{\rm F}^\dagger({\bm r})
\psi_{\rm F}({\bm r})
\psi_{\rm B}({\bm r}),
\label{eq.B1}
\end{eqnarray}
where $\psi_{\rm s=B,F}^\dagger({\bm r})$ are the Bose (s=B) and Fermi (s=F) field operators. To discuss the Hugenholtz-Pines condition, we add the following fictious Hamiltonian to the system:
\begin{equation}
H'(\lambda_{\rm B})=\int d{\bm r}{\hat \Psi}_{\rm B}^\dagger({\bm r})
{\hat \Lambda}_{\rm B},
\label{eq.B2}
\end{equation}
where 
\begin{eqnarray}
{\hat \Psi}_{\rm B}({\bm r})=
\left(
\begin{array}{c}
\psi_{\rm B}({\bm r}) \\
\psi_{\rm B}^\dagger({\bm r})
\end{array}
\right),
\label{eq.B3}
\end{eqnarray}
\begin{eqnarray}
{\hat \Lambda}_{\rm B}=
\left(
\begin{array}{c}
\lambda_{\rm B} \\
\lambda_{\rm B}^*
\end{array}
\right),
\label{eq.B4}
\end{eqnarray}
with $\lambda_{\rm B}$ being a complex number which is taken to be zero in the final stage of our discussion. $H'(\lambda_{\rm B})$ in Eq. (\ref{eq.B2}) determines the phase of the uniform Bose superfluid order parameter $\Delta_{\rm B}=\langle\psi_{\rm B}({\bm r})\rangle$ in the BEC phase. In particular, $\Delta_{\rm B}$ becomes real ($\equiv\Delta_{\rm B}^0$), when $\lambda_{\rm B}$ is chosen to be real ($\lambda_{\rm B}=\lambda_{\rm B}^*\equiv\lambda_B^0$).
\par
When we slightly modify $\lambda_{\rm B}$ in Eq. (\ref{eq.B4}) from $\lambda_{\rm B}^0$ as $\lambda_{\rm B}=\lambda_{\rm B}^0e^{i\delta\phi}$~($\delta\phi\ll 1$), while the Hamiltonian $H+H'$ is restored to the original form by the transformation $\psi_{\rm B}({\bm r})\to\psi_{\rm B}({\bm r})e^{i\delta\phi}$, the BEC order parameter $\Delta_{\rm B}=\langle\psi_{\rm B}({\bm r})\rangle$ and its conjugate are modified as,
\begin{eqnarray}
\left(
\begin{array}{c}
\Delta_{\rm B} \\
\Delta_{\rm B}^* 
\end{array}
\right)
&=&e^{i\delta\phi\tau_z}
\left(
\begin{array}{c}
\Delta_{\rm B}^0 \\
\Delta_{\rm B}^0
\end{array}
\right)
\simeq
\left(
\begin{array}{c}
\Delta_{\rm B}^0 \\
\Delta_{\rm B}^0
\end{array}
\right)
+\delta{\hat \Delta}_{\rm B},
\label{eq.B5}
\end{eqnarray}
where $\tau_z$ is the Pauli matrix, and
\begin{eqnarray}
\delta{\hat \Delta}_{\rm B}\equiv
i\delta\phi\Delta_{\rm B}^0\tau_z
\left(
\begin{array}{c}
1 \\
1
\end{array}
\right).
\label{eq.B6}
\end{eqnarray}
\par
Equation (\ref{eq.B6}) can also be obtained by evaluating the response of the BEC order parameter $\Delta_{\rm B}$ to the perturbation, 
\begin{eqnarray}
\delta H'\equiv H'(\lambda_{\rm B}^0e^{i\delta\phi})-H'(\lambda_{\rm B}^0)=
i\lambda_{\rm B}^0\delta\phi \int d{\bm r}
{\hat \Psi}_{\rm B}^\dagger({\bm r})\tau_z
\left(
\begin{array}{c}
1 \\
1
\end{array}
\right).
\label{eq.B7}
\end{eqnarray}
Using the standard linear response theory, we obtain\cite{Ry}
\begin{eqnarray}
\delta{\hat \Delta}_{\rm B}=i\lambda_{\rm B}^0\delta\phi{\hat G}_{\rm B}({\bm p}=0,i\omega^{\rm B}_n=0)\tau_z
\left(
\begin{array}{c}
1\\
1
\end{array}
\right),
\label{eq.B8}
\end{eqnarray}
where 
\begin{equation}
{\hat G}_{\rm B}({\bm p},i\omega_n^{\rm B})=
{1 \over i\omega^{\rm B}_n\tau_z-\xi_{\bm p}^{\rm B}-
{\hat \Sigma}_{\rm B}({\bm p},i\omega^{\rm B}_n)}
\label{eq.B9}
\end{equation}
is the $2\times 2$-matrix single-particle thermal Bose Green's function, with ${\hat \Sigma}_{\rm B}$ being the $2\times 2$-matrix self-energy. Equations (\ref{eq.B6}) and (\ref{eq.B8}) give
\begin{eqnarray}
\lambda_{\rm B}^0\tau_z
\left(
\begin{array}{c}
1\\
1
\end{array}
\right)
&=&
{\hat G}_{\rm B}(0,0)^{-1}\tau_z
\left(
\begin{array}{c}
1\\
1
\end{array}
\right)\Delta_{\rm B}^0
\nonumber
\\
&=&
\left[
\mu_{\rm B}-{\hat \Sigma}_{\rm B}(0,0)
\right]
\tau_z
\left(
\begin{array}{c}
1\\
1
\end{array}
\right)\Delta_{\rm B}^0.
\label{eq.B10}
\end{eqnarray}
Taking the limit $\lambda_{\rm B}^0\to 0$ in the BEC phase ($\Delta_{\rm B}^0\ne 0$), we find that Eq. (\ref{eq.B10}) is satisfied when
\begin{equation}
[\mu_{\rm B}
-{\hat \Sigma}_{\rm B}(0,0)]
\tau_z
\left(
\begin{array}{c}
1\\
1
\end{array}
\right)
=0.
\label{eq.B11}
\end{equation}
This is just the same form as the Hugenholtz-Pines condition\cite{Pines}. At $T_{\rm BEC}$, the off-diagonal components of the matrix self-energy ${\hat \Sigma}_{\rm B}(0,0)$ vanishes, so that the (1,1)-component of Eq. (\ref{eq.B11}) gives $\mu_{\rm B}-\Sigma_{\rm B}^{11}(0,0)=0$. Noting that $\Sigma_{\rm B}^{11}({\bm p},i\omega_n^{\rm B})$ equals $\Sigma_{\rm B}({\bm p},i\omega_n^{\rm B})$ appearing in Eq. (\ref{eq.8}) at $T_{\rm BEC}$, we obtain Eq. (\ref{eq.HP}).
\par
\par
\section{Derivation of Eqs. (\ref{eq.15}) and (\ref{eq.16})}
\par
The number equations (\ref{eq.15}) and (\ref{eq.16}) are obtained from the second-order virial expansion of the thermodynamic potential $\Omega$\cite{Liu},
\begin{equation}
\Omega=-{TV \over \lambda_T^3}
\left[
B_{10}z_{\rm F}+B_{20}z_{\rm F}^2+B_{01}z_{\rm B}+B_{02}z_{\rm B}^2+B_{11}z_{\rm F}z_{\rm B}
\right],
\label{ap1}
\end{equation}
by using the identity $N_{\rm s}=-(\partial \Omega/\partial \mu_{\rm s})_{\mu_{-{\rm s}}}$. Among the coefficients $B_{nn'}$ in Eq. (\ref{ap1}), $B_{10}$ and $B_{20}$ are actually the same as the corresponding virial coefficients in the case of a free Fermi gas\cite{Huang},
\begin{equation}
\Omega_{\rm F}=-T\sum_{\bm p}
\left[
1+e^{-\xi_{\bm p}^{\rm F}/T}
\right]
=-{TV \over \lambda_T^3}\sum_{n=1}^\infty{(-1)^{n+1} \over n^{5/2}}z_{\rm F}^n,
\label{ap2}
\end{equation}
so that one immediately finds $B_{n0}=(-1)^{n+1}n^{-5/2}$. In the same manner, $B_{01}$ and $B_{02}$ are obtained from the virial expansion in the case of a free Bose gas\cite{Huang},
\begin{equation}
\Omega_{\rm B}=T\sum_{\bm p}
\left[
1-e^{-\xi_{\bm p}^{\rm B}/T}
\right]
=-{TV \over \lambda_T^3}\sum_{n=1}^\infty{1 \over n^{5/2}}z_{\rm B}^n,
\label{ap3}
\end{equation}
that is, $B_{0n}=n^{-5/2}$.
\par
To determine $B_{11}$, it is convenient to substitute the expression for the second-order virial expansion of the grand partition function, 
\begin{equation}
\Xi=1+Q_{10}z_{\rm F}+Q_{20}z_{\rm F}^2+Q_{01}z_{\rm B}+Q_{02}z_{\rm B}^2+Q_{11}z_{\rm F}z_{\rm B},
\label{ap4}
\end{equation}
into $\Omega=-T\ln\Xi$, which gives 
\begin{equation}
B_{11}={\lambda_T^3 \over V}[Q_{11}-Q_{10}Q_{01}].
\label{ap5}
\end{equation}
Here, $Q_{nn'}={\rm Tr}_{nn'}[e^{-{\bar H}/T}]$ is a canonical partition function with $n$ fermions and $n'$ bosons, where ${\bar H}$ is given by Eq. (\ref{eq.1}) with $\xi_{\bm p}^{\rm s}$ being replaced by $\varepsilon_{\bm p}={\bm p}^2/(2m)$. Equation (\ref{ap5}) indicates that $B_{11}$ is just the difference of the two-particle canonical partition function between the interacting case ($Q_{11}$) and the non-interacting case ($Q_{10}Q_{01}$). When we change the variables of two particles into the center of mass coordinate from the relative one, and take into account the contribution from both the $s$-wave scattering component of the latter, because of the contact type inter-species interaction in Eq. (\ref{eq.1}) as well as from the bound state, we can rewrite Eq. (\ref{ap5}) as
\begin{align}
B_{11}
&
=
{\lambda_T^3 \over V}
\sum_{\bm K}e^{-{\bm K}^2/(4mT)}
\Bigg[
\bigg[
\sum_{k}e^{-k^2/(mT)}-\sum_{q}e^{-q^2/(mT)}
\bigg]
\nonumber
\\
&
+e^{{\varepsilon}_b/T}\theta({a_{\rm BF})}
\Bigg],
\label{ap5b}
\end{align}
where $\varepsilon_b=-1/(ma_{\rm BF}^2)<0$ is the $s$-wave bound state energy which exists only when $a_{\rm BF}^{-1}\geq 0$. Eq. (\ref{ap5b}) can again be written as,
\begin{equation}
B_{11}
=
2^{3/2}
\left[
\int_0^\infty dp
\left[\rho(p)-\rho_0(p)\right]
e^{-p^2/(mT)}
+e^{{\varepsilon}_b/T}\theta({a_{\rm BF})}
\right],
\label{ap6}
\end{equation}
where $\rho(p)$ ($\rho_0(p)$) is the momentum-space density of states in the interacting (non-interacting) case. Since the wave-function in the interacting case behaves as,
\begin{equation}
\Psi(r)\sim {1 \over r}\sin(pr+\delta_s(p)),
\label{ap7}
\end{equation}
(where $\delta_s$ is the phase shift associated with the inter-species interaction in Eq. (\ref{eq.1})), the interval $\Delta p$ of the quantum states in momentum space is given by
\begin{equation}
\Delta p=
{\pi \over \displaystyle R+{\partial\delta_s(p) \over \partial p}},
\label{ap8}
\end{equation}
where $R$ is the system size. The density of states $\rho(p)$ is then obtained as $\rho(p)=\Delta p^{-1}$. The density of states $\rho_0(p)$ in the non-interacting case is also obtained from Eq. (\ref{ap8}) where the second term in the denominator is absent. Substituting these results into Eq. (\ref{ap6}), one has\cite{Huang}
\begin{equation}
B_{11}
=
{2^{3/2}}
\left[
{1 \over \pi}
\int_0^\infty dp
{\partial\delta_s(p) \over \partial p}
e^{-p^2/(mT)}
+e^{{\varepsilon}_b/T}\theta({a_{\rm BF})}
\right].
\label{ap9}
\end{equation}
Using the relation $\tan(\delta_s(p))=-pa_{\rm BF}$ between the phase shift $\delta_s(p)$ and the scattering length $a_{\rm BF}$ in Eq. (\ref{eq.0}), we can further rewrite Eq. (\ref{ap9}) as
\begin{equation}
B_{11}=
{2^{3/2}}
\left[
{1\over \pi}
\int_0^\infty dp
{-a_{\rm BF} \over 1+(pa_{\rm BF})^2}
e^{-p^2/(mT)}
+e^{{\varepsilon}_b/T}\theta({a_{\rm BF})}
\right].
\label{ap10}
\end{equation}
Taking the limit from the weak coupling side in Eq. (\ref{ap10}), we obtain $B_{11}$ for any arbitrary interaction strength $a_{\rm BF}^{-1}$ as
\begin{align}
B_{11}
&=
{2^{3/2}}
\Bigg[
{-{\rm sgn}(a_{\rm BF}) \over \pi}
\int_0^\infty dx
{1 \over 1+x^2}
e^{-x^2/(mT|a_{\rm BF}|^2)}
\nonumber
\\
&+e^{{\varepsilon}_b/T}\theta({a_{\rm BF}})
\Bigg]
\nonumber
\\
&=
{2^{3/2}}
\left[
{-{\rm sgn}(a_{\rm BF}) \over 2}
[{1-{\rm erf}(y)}]
e^{y^2}
+e^{{\varepsilon}_b/T}\theta({a_{\rm BF})}
\right].
\label{ap11}
\end{align}
Here, ${\rm erf}(y)$ is the error function and $y=1/\sqrt{(mT|a_{\rm BF}|^2)}$. Taking the unitarity limit ($a_{\rm BF}^{-1}= \pm 0$) in Eq. (\ref{ap11}), we obtain $B_{11}$ as
\begin{equation}
B_{11}
=2^{3/2}\left(-{1\over 2}+1\right)
=\sqrt{2},
\label{ap12}
\end{equation}
where we have used $\varepsilon_b=0$, ${\rm erf}(y)=0$, and $e^{y^2}=1$, at the unitarity.
\par
\par

\end{document}